\documentclass[floats,floatfix,amssymb,prd,twocolumn,superscriptaddress,nofootinbib]{revtex4-2}

\usepackage[english]{babel}
\usepackage[utf8]{inputenc}
\usepackage{amsmath}
\usepackage{mathbbol}
\usepackage{amssymb}
\usepackage{bbold}
\usepackage{graphicx,amsfonts}
\usepackage{epsfig}
\usepackage[colorlinks=true,
linkcolor=blue,
urlcolor=blue,
citecolor=blue]{hyperref}
\usepackage{bm}
\usepackage{mathrsfs}
\usepackage{mathtools}
\usepackage{enumerate}
\usepackage{amsthm}
\usepackage{bbm}
\usepackage{comment}
\usepackage{physics}
\usepackage{url}
\usepackage{upgreek}
\usepackage[svgnames]{xcolor}
\usepackage[title]{appendix}

\begin{document}

\title{Ringdown stability: greybody factors as stable gravitational-wave observables}
\author{Romeo Felice Rosato}
\affiliation{Dipartimento di Fisica, Sapienza Università di Roma \& INFN, Sezione di Roma, Piazzale Aldo Moro 5, 00185, Roma, Italy}

\author{Kyriakos Destounis}
\affiliation{Dipartimento di Fisica, Sapienza Università di Roma \& INFN, Sezione di Roma, Piazzale Aldo Moro 5, 00185, Roma, Italy}

\author{Paolo Pani}
\affiliation{Dipartimento di Fisica, Sapienza Università di Roma \& INFN, Sezione di Roma, Piazzale Aldo Moro 5, 00185, Roma, Italy}

\begin{abstract}
The quasinormal mode spectrum of black holes plays a crucial role in the modelling of post-merger ringdown signals. However, the spectrum is extremely sensitive to small deformations of the system and describes the linear response only in a certain (not precisely defined) timeframe after the merger. We argue here that the greybody factors, recently shown to describe the ringdown spectral amplitude at relatively high frequencies, are instead stable under small perturbations of the system and free of certain ambiguities that plague the quasinormal mode spectrum. Our analysis also unveils a nontrivial interplay: while certain ringdown quantities are dominated by the contribution of spectrally unstable quasinormal modes, these modes conspire to produce stable observables. Thus, we propose a complementary approach to ringdown studies, which circumvents some limitations of the standard quasinormal mode description.
\end{abstract}

\maketitle

{\bf Introduction.} 
The black hole~(BH) spectroscopy program \cite{Dreyer:2003bv,Detweiler:1980gk,Berti:2005ys,Gossan:2011ha} is a cornerstone of strong-field tests of General Relativity~(GR)~\cite{LIGOScientific:2021sio,Berti:2015itd,Berti:2018vdi} and a unique way to test the nature of compact remnants formed after a coalescence~\cite{Cardoso:2019rvt}. It aims at extracting the remnant's vibrational spectrum, i.e. its  quasinormal modes~(QNMs)~\cite{Vishveshwara:1970zz,Kokkotas:1999bd,Berti:2009kk,Konoplya:2011qq}, during the ringdown stage of a binary merger. Within linear perturbation theory, the signal at intermediate times after the merger is described by a superposition of the QNMs~\cite{Leaver:1986gd} of the remnant till it relaxes to a stable end-state. If the remnant is a BH, GR predicts that the infinite tower of QNMs is uniquely described by its mass and spin, allowing for multiple null-hypothesis tests of gravity~\cite{Isi:2019aib,Franchini:2023eda}, the nature of the remnant~\cite{Maggio:2020jml,Maggio:2021ans,Maggio:2023fwy}, and the astrophysical environment around compact objects~\cite{Barausse:2014tra,Cardoso:2021wlq,Cardoso:2022whc,Destounis:2022obl}.

While QNMs are the gold standard to perform BH spectroscopy, in recent years a growing amount of subtleties related to them have emerged \cite{Baibhav:2023clw}. Firstly, QNMs do not form a complete set of solutions (eigenfunctions) since they are the eigenvalues of a non-selfadjoint (non-Hermitian) system~\cite{Kokkotas:1999bd}. This implies that a generic, small perturbation cannot be decomposed only in QNMs; indeed, the ringdown also displays an early-time response that depends on the details of the initial merger conditions, and a late-time power-law tail~\cite{Price:1972pw,Gundlach:1993tp,Barack:1998bw} due to back-scattering off the effective potential. The exact start of the intermediate stage governed by the QNMs is an open problem and might not have a clear-cut answer~\cite{Giesler:2019uxc,Cotesta:2022pci,Isi:2022mhy,Carullo:2023gtf,Isi:2023nif}. This is an issue already within linear perturbation theory, leaving aside the fact that GR is intrinsically nonlinear \cite{Gleiser:1995gx,Gleiser:1998rw,Ioka:2007ak,Nakano:2007cj,Brizuela:2009qd,Pazos:2010xf,Ripley:2020xby,Loutrel:2020wbw}, understanding when the nonlinear stage transits toward a perturbative regime is still highly debated~\cite{Sberna:2021eui,Cheung:2022rbm,Mitman:2022qdl,Kehagias:2023ctr,Perrone:2023jzq,Cheung:2023vki,Redondo-Yuste:2023ipg,Redondo-Yuste:2023seq,Yi:2024elj,Zhu:2024dyl,Zhu:2024rej}. In this context, the role of QNM overtones, nonlinearities, and ringdown starting time are outstanding interconnected open questions~\cite{Baibhav:2023clw}.

Another concern regarding QNMs is that they are extremely sensitive to small perturbations of the system~\cite{Nollert:1996rf,Daghigh:2020jyk,Jaramillo:2020tuu} either in terms of deformations of the background or of the boundary conditions. This implies that the QNM spectrum of a BH might be drastically altered in the presence of environmental effects~\cite{Barausse:2014tra,Barausse:2014pra,Cheung:2021bol,Berti:2022xfj} or of any form of near-horizon structure \cite{Cardoso:2016rao,Cardoso:2016oxy,Cardoso:2017cqb,Abedi:2020ujo}, although the prompt ringdown phase in time domain is much less affected~\cite{Cardoso:2016rao,Cardoso:2016oxy,Cardoso:2017cqb,Mirbabayi:2018mdm,Berti:2022xfj,Kyutoku:2022gbr}. Nevertheless, Ref. \cite{Jaramillo:2021tmt} proved that a QNM spectral instability is indeed accessible in the time-domain gravitational-wave signal, though a large signal-to-noise ratio will be required and such a detection will pose a challenging data analysis problem.

While all these issues might not necessarily be an obstacle for an actual realization of the BH spectroscopy program --~at least within the accuracy of current detectors~-- they unveil that the ``vanilla'' extraction of QNMs from gravitational-wave signals is much more subtle than historically expected. Originally~\cite{Detweiler:1980gk,Dreyer:2003bv,Berti:2005ys,Gossan:2011ha}, BH spectroscopy was presented as a particularly simple and clean way to test gravity and the nature of remnants, but the emergence of these issues suggests at least a more cautious view.

Here we propose a complementary approach that circumvents some of the above subtleties while elucidating the physical interpretation of the QNM spectral instability~\cite{Jaramillo:2020tuu,Destounis:2021lum,Gasperin:2021kfv,Boyanov:2022ark,Jaramillo:2022kuv,Sarkar:2023rhp,Destounis:2023nmb,Arean:2023ejh,Cownden:2023dam,Destounis:2023ruj,Courty:2023rxk,Boyanov:2023qqf,Cao:2024oud,Cardoso:2024mrw}. We show that the BH greybody factors~(GFs) --~functions characterizing the tunnelling probability of perturbations through the BH effective potential~\cite{Hawking:1975vcx}~-- are stable under small deformations of the background and are associated with quantities that can be obtained by a superposition of QNMs, despite the latter being spectrally unstable. Thus, our analysis unveils a remarkable interplay: spectrally unstable QNMs conspire to produce stable observables. An analogous result was recently obtained for the scattering cross-section using (spectrally unstable) Regge poles~\cite{Torres:2023nqg}. Moreover, the stability of wave scattering amplitudes was discussed in analogy with the phase shift of the $S$-matrix in a quantum system~\cite{Kyutoku:2022gbr}.

Recently, the GFs were shown to describe the spectral amplitude of the ringdown signal at frequencies higher than that of the fundamental QNM~\cite{Oshita:2022pkc,Oshita:2023cjz,Okabayashi:2024qbz}. Our results, together with~\cite{Oshita:2022pkc,Oshita:2023cjz,Okabayashi:2024qbz}, suggest a route to a complementary study of the BH ringdown using GFs; quantities that form stable ringdown observables and evade some debatable aspects that QNMs exhibit. Henceforth we use $G=c=1$ units.

{\bf Stability of BH GFs.}
We consider perturbations of a spherically-symmetric BH in the Regge-Wheeler-Zerilli formalism~\cite{Regge:1957td,Zerilli:1970se}, although our analysis can be straightforwardly extended to rotating spacetimes using Teukolsky's formalism~\cite{Teukolsky:1973ha,Press:1973zz}.

Within linear perturbation theory, the BH response to an external perturbation in the frequency domain is described by a one-dimensional radial equation~\cite{Zerilli:1970se,SASAKI198185}
\begin{equation}\label{Regge-Zerilli_source}
 \Bigg[{d^2 \over dr_*^2} + \omega^2 - V_l(r)\Bigg] X_{lm\omega}=S_{lm\omega}(r)
\end{equation}
where $r_*$ is the tortoise coordinate, $\omega$ is the frequency, and $(l,m)$ are the spherical-harmonic indices. The effective potential $V_l$ and the source $S_{lm\omega}$ (the latter being related either to the stress-energy tensor of the perturbation in a Fourier decomposition~\cite{Zerilli:1970se} or to the initial data of $X_{lm\omega}$ in a Laplace decomposition~\cite{Leaver:1986gd}) are different for axial and polar perturbations. 

First, we consider the homogeneous equation. The BH GF is the transmission coefficient of a scattering problem identified by the boundary conditions
\begin{equation}\label{boundary_refl/trasm}
{X}_{lm\omega}= \begin{cases}
     e^{- \imath \omega r_*} \,\,\,r_*\to-\infty \\
    A^{\rm in}_{lm\omega} e^{- \imath \omega r_*}+ A^{\rm out}_{lm\omega} e^{+ \imath \omega r_*}\,\,\, r_*\to+\infty
 \end{cases}\,.
\end{equation}
In turn, QNMs are the complex frequencies satisfying the above conditions with $A^{\rm in}_{lm\omega}=0$. Due to scattering off the potential barrier, we can define the reflectivity and transmissivity of the background spacetime as (see, e.g.,~\cite{Brito:2015oca}):
\begin{equation}
  \mathcal{R}_{lm}(\omega)=\Bigg|{A^{\rm out}_{lm\omega} \over A^{\rm in}_{lm\omega}}\Bigg|^2\,, \qquad \Gamma_{lm}(\omega)=\Bigg|{1 \over A^{\rm in}_{lm\omega}}\Bigg|^2\,, \label{RandT}
\end{equation}
where $\Gamma_{lm}$ is the GF and energy conservation enforces $\mathcal{R}_{lm}+\Gamma_{lm}=1$. This quantity plays a pivotal role in the linear response of a BH, being associated with the BH absorption cross-section, the rate of Hawking evaporation~\cite{Hawking:1975vcx}, and the ringdown spectral amplitude. For a Kerr BH, the GF for a given $(l,m)$ is a function of $\omega$ that, just like QNMs, depends only on the mass and spin. 

\begin{figure}[t]
    \centering
    \includegraphics[width=0.5\textwidth]{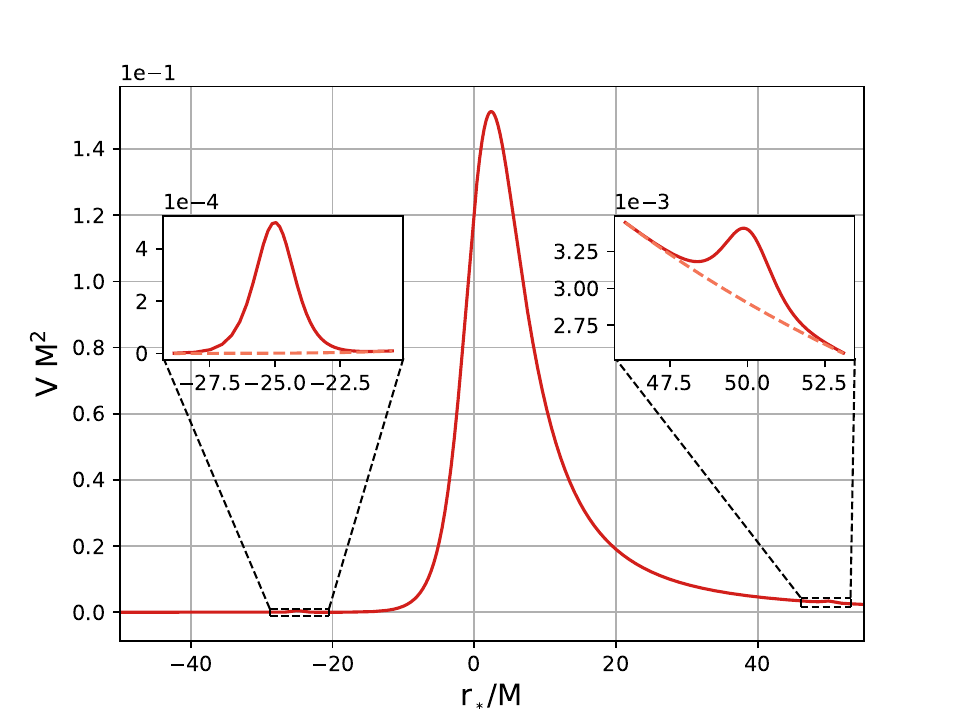}
    \caption{Perturbed effective potential, Eq.~\eqref{bump}, for $l=2$, $\epsilon=5\times 10^{-4}$, and two cases for the bump location, i.e. $c=-25M$ (left inset) and $c=50M$ (right inset).}
    \label{fig:potential}
\end{figure}

To study the stability of GFs against small perturbations of the system, we consider an infinitesimal P\"oschl-Teller bump added to the original potential~\cite{Cheung:2021bol}
\begin{equation}\label{bump}
   V^\epsilon_l=
   \left(1-\frac{2M}{r}\right)\left(\frac{l(l+1)}{r^2}-\frac{6M}{r^3}\right)
   +\frac{\epsilon}{M^2}\sech^2\left[\frac{r_*-c}{M}\right],
\end{equation}
where $c$ and $\epsilon\ll1$ parametrize the location and amplitude of the bump, respectively (see Fig.~\ref{fig:potential}). When $c\gg M$ (with $M$ being the BH mass), the bump parametrizes distant perturbations, such as environmental effects~\cite{Barausse:2014pra,Cheung:2021bol}. When $c<0$ but $|c|\gg M$, the bump parametrizes local perturbations, such as near-horizon structures predicted by quantum-gravity models~\cite{Cardoso:2017cqb,Bena:2022rna}, plasma \cite{Cardoso:2020nst,Cannizzaro:2024yee} and dark-matter overdensities \cite{Bertone:2024wbn} accumulated near the horizon. We will consider $c$ as a free parameter, showing that the phenomenology is similar whenever $|c|\gg M$. Note that we simply quantify the ``smallness'' of the perturbing bump through the parameter $\epsilon$, as done in other work~\cite{Cheung:2021bol,Torres:2023nqg}, A more mathematically-robust way to measure the amplitude of the perturbation is through the energy norm \cite{Trefethen:1993,Driscoll:1996,Jaramillo:2020tuu,Jaramillo:2021tmt,Gasperin:2021kfv} that will eventually not take into account only the bump amplitude $\epsilon$, but also the bump location, $c/M$. Indeed, the energy norm of a similar type of bump, that has a Gaussian distribution form, has very recently been established \cite{Boyanov:2024fgc}. In this example of an \emph{ad-hoc} addition of a Gaussian bump to the Regge-Wheeler potential, there is a clear correlation between the energy norm of the external perturbation and the location of the bump, thus assuming a tiny $\epsilon$ does not necessarily means a small perturbation overall. Nevertheless, in the following, we will present examples of P\"oschl-Teller bumps of amplitude $\epsilon\ll 1$ placed at distances $c/M=50,\,100,\,250$ from the BH potential and the GFs still remain stable. In a sense, our results are actually conservative, since, even if the energy norm is large, we find that the GFs are stable, so they will most likely remain stable also for perturbations with smaller norms. Hence, we argue that our results should be general, for small or large perturbation norms, and should not directly depend on the specific magnitude or shape of the perturbation, though a universal proof is still lacking.

\begin{figure*}[ht]
    \centering
    \includegraphics[width=0.95\textwidth]{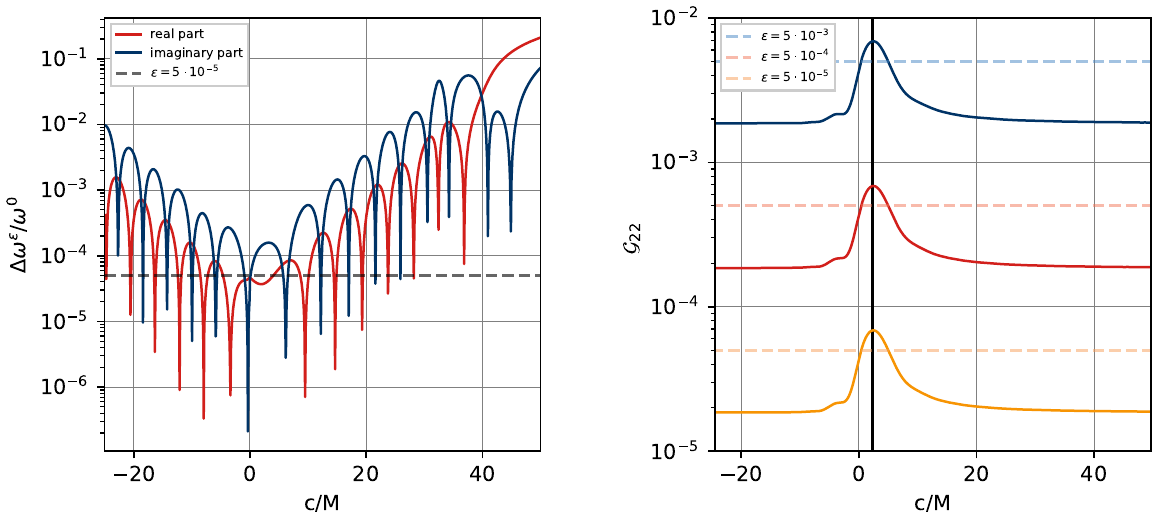}
    \caption{Comparison between the spectral instability of the fundamental QNM (left panel) and the stability of the (integrated) GF (right panel) of a Schwarzschild BH under a small perturbation centred at $r_*=c$ with amplitude $\epsilon\ll1$. On the left we fix $\epsilon=5\times 10^{-5}$ (black dashed line) while on the right we consider three values of $\epsilon$ (denoted by dashed lines). The value of $\mathcal{G}_{lm}$ is maximized when $c$ corresponds to the peak of the unperturbed potential, whose value is indicated by a vertical solid line.}
    \label{SchwPerturb}
\end{figure*}

In Fig.~\ref{SchwPerturb} we compare the spectral instability of the fundamental QNM~\cite{Barausse:2014pra,Jaramillo:2020tuu,Cheung:2021bol} (left panel) with the stability of the GFs (right panel) for varying $c/M$ and perturbation scale $\epsilon$. For the QNMs, we show the relative difference  $\Delta\omega^\epsilon_{R,I}/\omega^0_{R,I}=|\omega^\epsilon_{R,I}(c)/\omega^0_{R,I}-1|$, where $\omega^\epsilon_{R,I}(c)$ and $\omega^0_{R,I}$ are the (real and imaginary part of the) QNMs of the perturbed and unperturbed system, respectively. These were computed using both direct integration~\cite{Pani:2013pma} and continued fractions adapted from~\cite{LeaverCFextension:PhysRevD.41.2986,Dolanetal:PhysRevD.101.104035} (to allow for a potential that is not polynomial in $M/r$), finding perfect agreement. Since the GF is a function of the frequency, on the right panel we plot the integrated quantity
\begin{equation}\label{Deltal}
    \mathcal{G}_{lm}={ \int^\infty_{0} \Big|\Gamma^\epsilon_{lm} (\omega,c)-\Gamma_{lm} (\omega)\Big| d\omega \over \int^\infty_{0} \Gamma_{lm} (\omega) d\omega }\,,
\end{equation}
where the unperturbed GFs, $\Gamma_{lm}(\omega)$, and perturbed ones, $\Gamma^\epsilon_{lm} (\omega,c)$, have been computed through direct integration using analytical high-order series expansions to reach high precision~\cite{Pani:2013pma,Brito:2015oca}. The above dimensionless quantity $\mathcal{G}_{lm}$ is positive definite and quantifies how the GF (for a fixed value of $l$) behaves under the perturbation in Eq.~\eqref{bump} for any frequency.

As it is clear from Fig.~\ref{SchwPerturb}, the QNM deviation from the unperturbed system grows exponentially with $|c|$, so that eventually $\Delta \omega^\epsilon / \omega^0 \gg \epsilon$, in agreement with previous results~\cite{Barausse:2014pra,Jaramillo:2020tuu,Cheung:2021bol,Cardoso:2024mrw} and as expected for a spectrally unstable quantity~\cite{Jaramillo:2020tuu,Cheung:2021bol}. However, the GFs display a completely different behavior: they are bounded and remain ${\cal O}(\epsilon)$, as expected for a quantity that is stable under small perturbations. In fact, at variance with the QNM case, the stability of the GF improves when $|c|$ increases, and $\mathcal{G}_{lm}<\epsilon$ as $|c|\to \infty$. Interestingly, $\mathcal{G}_{lm}$ has a maximum when $c$ corresponds to the peak of the unperturbed potential ($c\approx 2.39 M$, for the $l=2$ axial case), denoted by a vertical solid line.

\begin{figure*}[ht]
    \centering
    \includegraphics[width=0.95\textwidth]{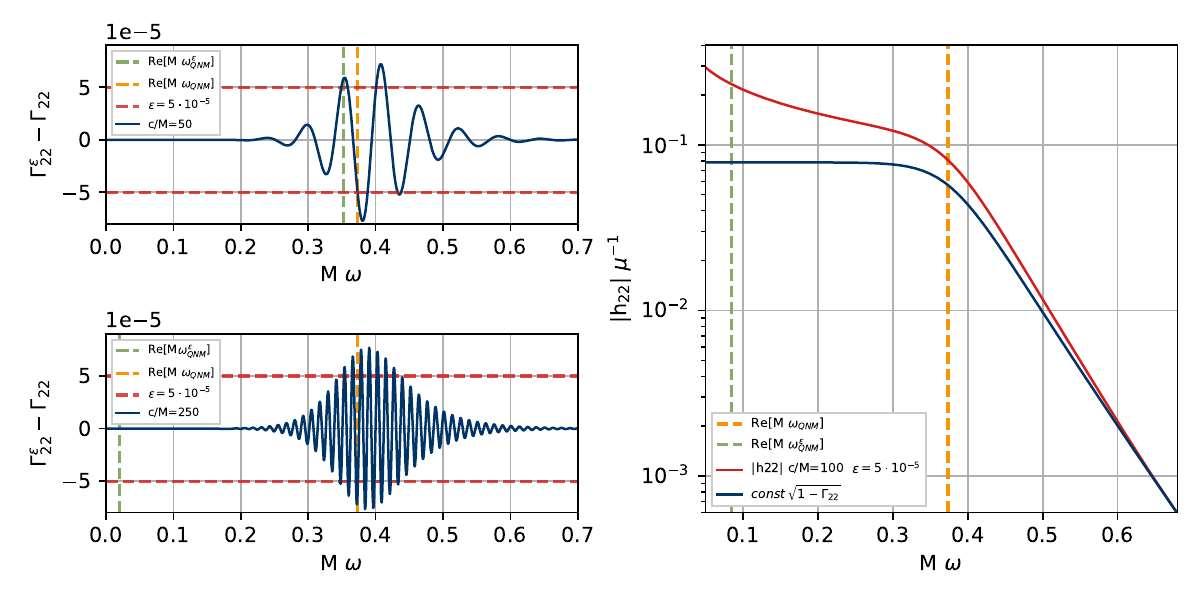}
    \caption{Left: Difference between the perturbed and unperturbed GF as a function of the frequency for some representative value of $\epsilon$ and $c$. The difference is always ${\cal O}(\epsilon)$ or smaller and peaks at the frequency of the \emph{unperturbed} fundamental QNM. Right: Gravitational-wave amplitude in the frequency domain for a point particle in radial infall with mass $\mu$ and $E=1, L=0$. We compare the signal with an \emph{analytical} model, $|h_{22}|\propto\sqrt{1-\Gamma_{22}}$. Being stable observables, the difference between $h_{22}$ or $\Gamma_{22}$ in the perturbed/unperturbed case is indistinguishable on the scale used. In both panels the vertical green/orange lines denote the frequency of the perturbed/unperturbed fundamental QNM.}
\label{fig:GammANDh}
\end{figure*}

We find this behavior for both the integrated GF in Eq.~\eqref{Deltal} and $\Gamma_{lm}(\omega)$ at any frequency, as shown in Fig.~\ref{fig:GammANDh}. This result is consistent with the analysis in~\cite{Kyutoku:2022gbr}, where it was shown that the main effect of the bump appears in the phase shift of the scattered wave rather than in its amplitude. The left panel of Fig.~\ref{fig:GammANDh} shows another remarkable feature. In the unperturbed case, the GF interpolates between $\Gamma_{lm}\sim 0$ at low frequencies and $\Gamma_{lm}\sim1$ at high frequencies with the smooth transition occurring at the fundamental QNM frequency (see~\cite{Bonelli:2021uvf} for an analytical expression of the GF at any frequency). The GF of the perturbed system displays exactly the same behavior, despite the fact that the QNMs of the perturbed system acquire ${\cal O}(1)$ corrections. This is consistent with the fact that the GF is more sensitive to local features of the effective potential near the maximum, so the transition occurs approximately at the frequency of the light ring of the background, which in the perturbed case does not correspond to that of the QNM~\cite{Cardoso:2016rao,Cardoso:2014sna}. This behavior appears for any $l$ and for both axial and polar perturbations.

Even though we should expect that the GF displays sharp Breit-Wigner resonances~\cite{Berti:2009wx,Mascher:2022pku} at the frequency of long-lived modes ($|\omega_I^\epsilon|\ll \omega_R^\epsilon$), we did not find any for this perturbed potential, probably because the quality factor of the perturbed QNMs remains moderate. 
For example, when $c=100M$ and $\epsilon=10^{-2}$, the unperturbed fundamental mode $\omega^0 M \approx 0.37-\imath 0.09$ migrates to $\omega^\epsilon  M\approx 0.052- \imath 0.009$ in the axial case, but the GF does not display any special feature at $\omega=\omega^\epsilon_R$. In any case, these resonances can possibly appear only at low frequencies~\cite{Cardoso:2014sna} and would not contaminate the large-frequency behavior of the GF. As discussed below and in~\cite{Oshita:2023cjz,Okabayashi:2024qbz}, the latter is directly connected to the ringdown amplitude.

{\bf GFs, QNMs, \& ringdown.}
Although we found that the GF is stable for \emph{any} frequency, it was recently argued that the spectral amplitude, $|h_{lm}(\omega)|\propto |X_{lm\omega}|$ (see Appendix for the exact definition), is well described by the reflectivity $\sqrt{1-\Gamma_{lm}
(\omega)}$ at frequencies larger than the fundamental QNM~\cite{Oshita:2023cjz,Okabayashi:2024qbz}. To further shed light on this connection, here we study the signal emitted by a point-particle orbiting a BH. The signal is described by Eq.~\eqref{Regge-Zerilli_source}, where the source depends on the specific energy $E$ and angular momentum $L$ of the particle~\cite{Zerilli:1970se,SASAKI198185,Martel:2005ir}. This equation can be solved with standard Green function methods. The $l=m=2$ spectral amplitude is shown in the right panel of Fig.~\ref{fig:GammANDh} for $E=1, L=0$ (see Appendix for further cases). We show the perturbed case for $c=100M$ and $\epsilon=5\times10^{-5}$, that perfectly overlaps with the unperturbed case. Interestingly, the arguments in \cite{Oshita:2023cjz,Okabayashi:2024qbz} apply to the ringdown both in the extreme-mass-ratio limit~\cite{Oshita:2023cjz} and in numerical-relativity simulations of comparable-mass binaries~\cite{Okabayashi:2024qbz}. We find the same behavior for the perturbed system, and also in this case the cross-over occurs at the frequency of the \emph{unperturbed} QNM. This shows that the ringdown spectral amplitude at $\omega\gtrsim {\rm Re}[\omega_{lm0}]$ is a stable observable quantity. Remarkably, this result relies on the fact that $|h_{lm}(\omega)|$ depends only on the absolute value of the signal, whereas individually its real and imaginary parts are sensitive to perturbations (see~\cite{Kyutoku:2022gbr} and Appendix).

\begin{figure}[ht]
    \centering
    \includegraphics[width=0.5\textwidth]{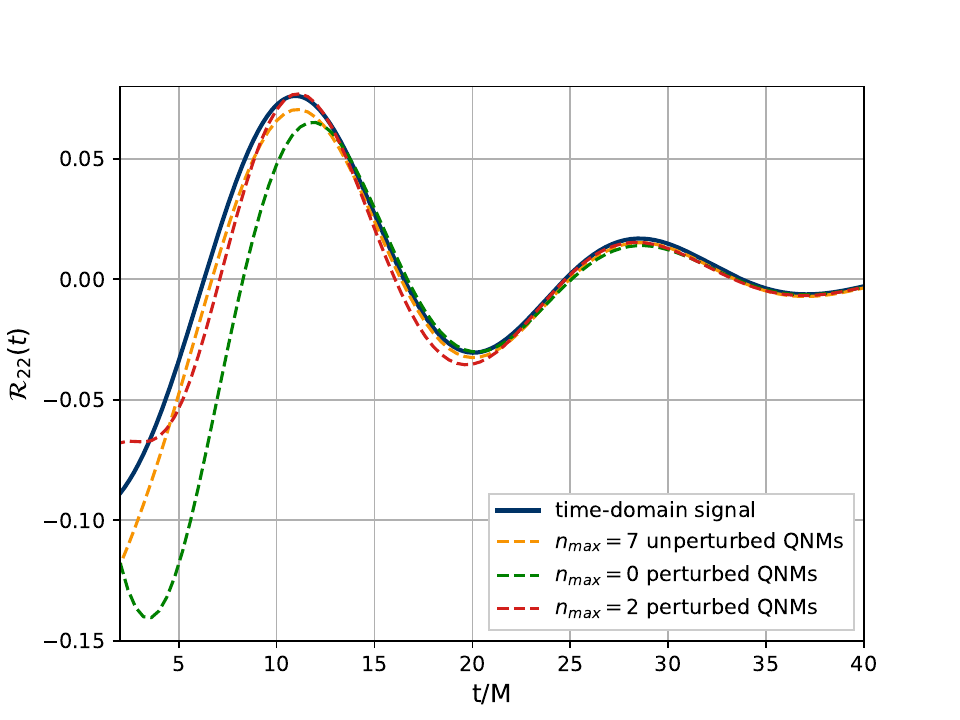}
    \caption{Fourier transform of the amplitude ratio, $A_{lmn\omega}^{\rm out}/A_{lmn\omega}^{\rm in}$, compared to its QNM decomposition in terms of excitation factors adding up to $n_{\rm max}$ modes (see Eq.~\eqref{QNMdecomposition}), for $c=15M$ and $\epsilon=10^{-3}$. The time-domain signal is stable under small perturbations and can be recovered by a superposition of \emph{either} unperturbed or perturbed QNMs, despite the spectral instability of the latter: here, $\omega^0_RM\approx(0.374,0.347,0.301)$,~$\omega^0_IM\approx-(0.089,0.274,0.478)$, and $\omega^\epsilon_RM\approx(0.374,0.335,0.489)$,~$\omega^\epsilon_IM\approx-(0.092,0.182,0.266)$, where~$n=(0,1,2)$.}
    \label{fig:QNMdecomp}
\end{figure}

Finally, it is instructive to compute the inverse Fourier transform of the amplitude ratio that gives the reflectivity and GF through Eq.~\eqref{RandT}, namely $R_{lm}(t)=\frac{1}{2\pi} \int_{-\infty}^{+\infty} d\omega \, \frac{A^{\rm out}_{lm\omega}}{A^{\rm in}_{lm\omega}} e^{-\imath\omega t}$. We can compute this either fully numerically or by performing a contour integral in the complex plane~\cite{Leaver:1986gd}. In the latter case the main contribution comes from the simple poles at the QNMs corresponding to the complex roots, $A^{\rm in}_{lm\omega}\sim \gamma_{lmn}(\omega-\omega_{lmn})$, where $n$ is the overtone number. This can be computed through the residue theorem as
\begin{equation}
    R_{lm}(t) \approx -2 {\rm Re}\left[\imath \sum_{n=0}^{n_{\rm max}}  \frac{ \left.A^{\rm out}_{lm\omega}\right|_{\omega=\omega_{lmn}}}{\gamma_{lmn}} e^{-\imath \omega_{lmn} t}\right]\,, \label{QNMdecomposition}
\end{equation}
where the factor $2$ accounts for the QNMs with opposite real part~\cite{Leaver:1986gd}, and the terms of the series are directly related to the BH excitation factors~\cite{Andersson:1995zk,Glampedakis:2003dn,Berti:2006wq,Zhang:2013ksa,Silva:2024ffz}, which are source-independent. In Fig.~\ref{fig:QNMdecomp}, we compare the full result for $R_{lm}(t)$ at intermediate times (where the perturbed and unperturbed cases are indistinguishable~\cite{Cardoso:2016oxy,Cardoso:2016rao,Cardoso:2017cqb,Berti:2022xfj}) with that obtained by summing over a certain number of QNMs, i.e., using Eq.~\eqref{QNMdecomposition} for different values of $n_{\rm max}$, both in the unperturbed and in the perturbed case. Remarkably, even if overtones display ${\cal O}(1)$ deviations in the perturbed case (see Appendix), their inclusion improves the recovery of the stable observable quantity at early times compared to the case of a single fundamental QNM, and in fact even with respect to the $n_{\rm max}=7$ unperturbed case. To the best of our knowledge this is the first time that the (stable) time-domain signal is shown to be obtained as a superposition of (spectrally-unstable) QNMs.

{\bf Discussion.}
Despite the spectral instability of BH QNMs, it is reassuring that other relevant quantities related to BH perturbation theory, such as the GFs and the reflectivities, are stable under small perturbations of the system. In addition, we find that the stability of the ringdown spectral amplitude, $|h_{lm}(\omega)|$, extends to other observable quantities, such as the emitted energy (see Appendix).

The connection between the (stable) GF and the ringdown unveils a number of interesting features in what concerns tests of gravity. Since the GF describes the ringdown amplitude at relatively high frequencies, we expect it to be more sensitive to short-lengthscale modifications in the strong-field regime compared to usual BH QNMs. This might be an advantage to test high-curvature corrections to GR as those predicted within effective field theories~\cite{Berti:2015itd,Cardoso:2018ptl,Cano:2023jbk}, and is generically complementary to ordinary BH QNM tests. In this respect, one might devise a null-hypothesis test of the Kerr metric using GFs with different $(l,m)$'s, along the lines of standard BH spectroscopy. In particular, $\Gamma_{22}(\omega)$ can be extracted from $|h_{22}|$ and can be used to infer the mass and spin of the remnant~\cite{Oshita:2023cjz,Okabayashi:2024qbz}. Then, any other multipole of $|h_{lm}|$ can be used to extract $\Gamma_{lm}(\omega)$, which is uniquely determined by the mass and spin if the remnant is a Kerr BH. Thus, measuring (say) $\Gamma_{22}(\omega)$ and $\Gamma_{33}(\omega)$ would provide a null test. At variance with the same test using QNMs, the one based on the GFs is not contaminated by spectral instabilities, as far as we have checked, nor by overtones (since the GFs depend only on $(l,m)$); it also contains more information, since a successful test should fit the entire functions $\Gamma_{lm}(\omega)$ and not only numbers. In the Appendix, we show that for a Schwarzschild BH in GR the exponential behavior of the reflectivity ${\cal R}_{lm}(\omega)=1-\Gamma_{lm}(\omega)$ at large frequencies is universal and independent of $l$ (in agreement with~\cite{Sanchez:1976fcl}), but this property might be different for a Kerr BH, in modified theories of gravity, or for alternative compact objects. We plan to perform a detailed examination of tests of gravity based on the GFs in future work, such as ringdown analyses with QNM filters constructed with the GFs of BH remnants \cite{Ma:2022wpv,Ma:2023vvr,Ma:2023cwe} that have been proven to devise stable data-analysis pipelines for gravitational-wave events.

Besides its direct connection to the ringdown, the GF is also associated to the absorption cross-section of a plane wave~\cite{Mashhoon:1973zz,Fabbri:1975sa,Sanchez:1977si,Andersson:1995vi} which implies that this quantity is also stable \cite{Torres:2023nqg}. Finally, the GF accounts for the differences between Hawking radiation and a perfect black-body spectrum \cite{Hawking:1975vcx}. The BH temperature is a local quantity related to the surface gravity at the horizon \cite{Hawking:1976de}, so it is also stable. The stability of $\Gamma_{lm}$ then ensures that the Hawking emission rate is also a stable quantity.

Finally, we measured the ``smallness'' of the external P\"oschl-Teller perturbation by its amplitude $\epsilon$ and not by its energy norm, that should also include the location of the bump. Nevertheless, even if energy-normalized perturbations are considered, as done recently~\cite{Boyanov:2024fgc}, the GFs should, in principle remain stable, in accord to our analysis where not only $\epsilon$ but also $c/M$ was extensively varied.

\textbf{Note:} After the submission of our manuscript, a manuscript appeared on the arXiv~\cite{Oshita:2024fzf}, with a very similar analysis. Their results agree very well with ours, thus confirm the two independent analyses.

\begin{acknowledgments}
We are indebted to Koutarou Kyutoku, Sizheng Ma, and Jaime Redondo Yuste for comments on the draft. We are also thankful to Vitor Cardoso and Naritaka Oshita for fruitful discussions. We acknowledge partial support by the MUR PRIN Grant 2020KR4KN2 ``String Theory as a bridge between Gauge Theories and Quantum Gravity'' and by the MUR FARE programme (GW-NEXT, CUP:~B84I20000100001). 
\end{acknowledgments}

\section*{Supplemental material}
In this Appendix we collect some further results that extend those presented in the main text.
\appendix
\section{Stability of the GFs}
\begin{figure}[ht]
	\centering
	\includegraphics[width=0.49\textwidth]{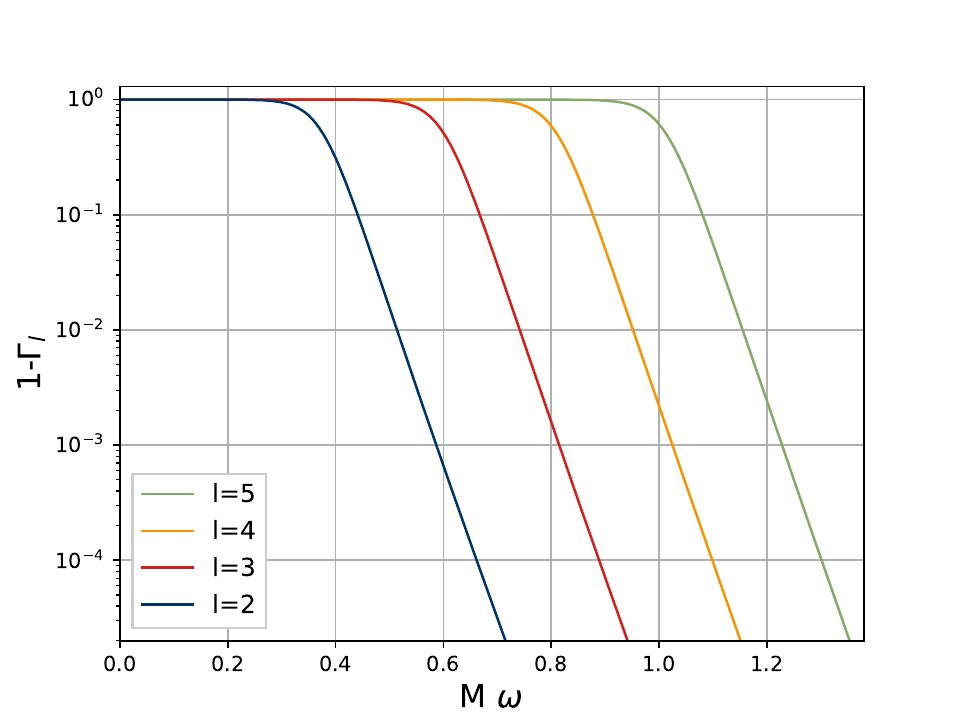}
	\caption{Reflectivity ${\cal R}_{lm}(\omega)=1-\Gamma_{lm}(\omega)$ of a Schwarzschild BH as a function of the frequency $\omega$ for different values of $l=2,3,4,5$. The GFs are stable so the difference between unperturbed and perturbed case is irrelevant if $\epsilon\ll1$. In particular, the exponential fall-off starts at frequencies corresponding to the unperturbed fundamental QNM frequency and is universal for any $l$.}
	\label{fig:variousgamma}
\end{figure}

Figure~\ref{fig:variousgamma} shows the reflectivity ${\cal R}_{lm}(\omega)=1-\Gamma_{lm}(\omega)$ of a Schwarzschild BH for $l=2,3,4,5$. In all these cases we found that the GFs are stable and the transition from $\Gamma_{lm}(\omega)\sim0$ to $\Gamma_{lm}(\omega)\sim1$ occurs at the frequency of the \emph{unperturbed} fundamental QNM frequency. In addition, as shown in Fig.~\ref{fig:variousgamma} the behavior at large frequencies is exponential, with the same slope, for all $l$.

\begin{figure}[t]
	\centering
	\includegraphics[width=0.48 \textwidth]{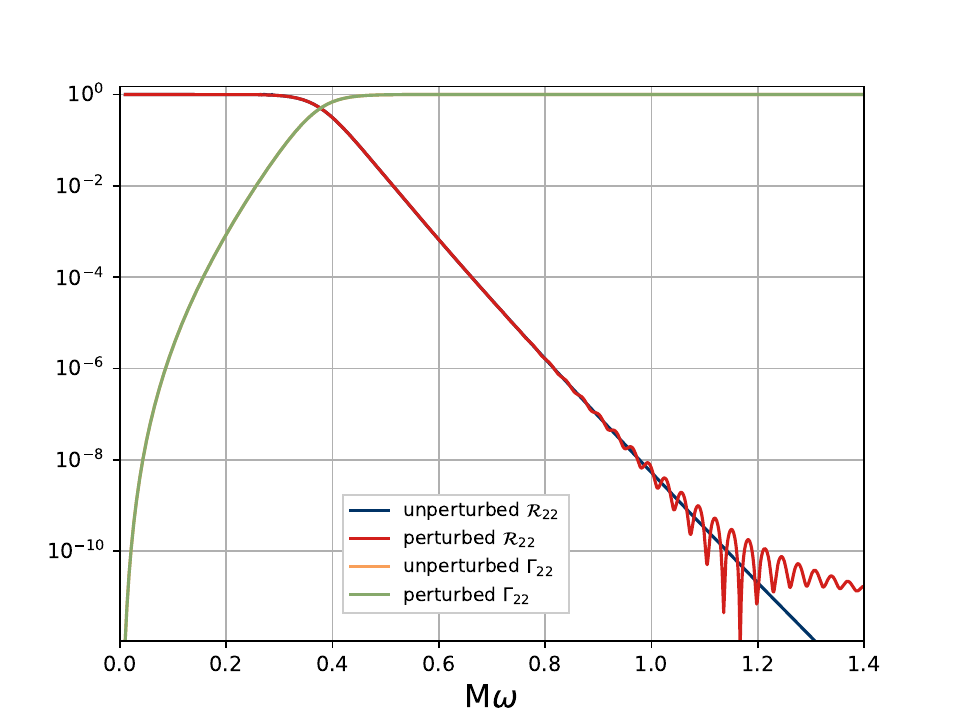}
	\caption{Reflectivity and GF for a perturbed Schwarzschild BH with $c=100M$ and $\epsilon=5 \times 10^{-4}$ compared to the unperturbed case. The two curves for $\Gamma_{22}$ are indistinguishable.}
	\label{fig:unstableR}
\end{figure}

Note that, despite $\Gamma_{lm}$ being stable, i.e. $\Delta \Gamma_{lm}/\Gamma_{lm}^0={\cal O}(\epsilon)$, the fact that the reflectivity is defined as ${\cal R}_{lm}=1-\Gamma_{lm}$ implies that $\frac{\Delta {\cal R}_{lm}}{{\cal R}_{lm}^0}=\frac{\Gamma_{lm}^0}{1-\Gamma_{lm}^0}{\cal O}(\epsilon)$, so that the relative difference of the reflectivity grows in the \emph{large-frequency} regime as $\Gamma_{lm}^0\to 1$ exponentially. This is simply due to the fact that ${\cal R}_{lm}$ is exponentially suppressed at large frequencies, so ${\cal O}(\epsilon)$ corrections are relatively big, while in that regime they are negligible for $\Gamma_{lm}\sim1$.
Indeed, as shown in Fig.~\ref{fig:unstableR}, this only occurs when ${\cal R}_{lm}$ is already exponentially suppressed, while the reflectivity is stable at smaller frequencies. Note that for sufficiently high frequencies ($M\omega > 1.2$), the perturbed ${\cal R}_{lm}$ exhibits a tail, which is also observed in \cite{Oshita:2024fzf}. It is important to emphasize that such a deviation from the unperturbed function is entirely expected, given that $\frac{\Delta R_{lm}}{R_{lm}^0} = \frac{\Gamma_{lm}^0}{1-\Gamma_{lm}^0}O(\epsilon)$. Nevertheless, this does not compromise the stability of the greybody factors.

\begin{figure}[t]
	\centering
	\includegraphics[width=0.48 \textwidth]{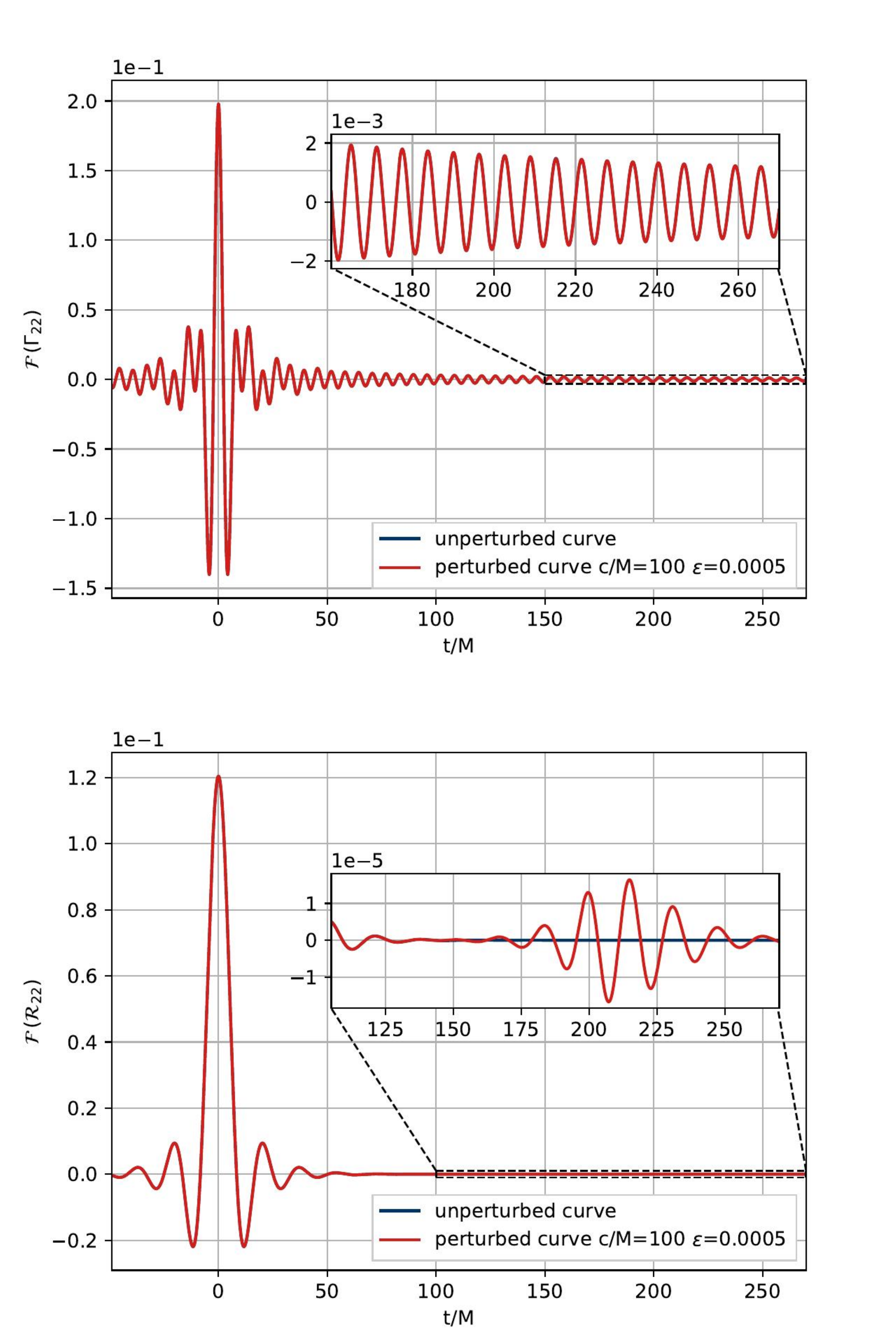}\\
	\caption{Fourier transform of $\Gamma_{22}$ (top panel) and ${\cal R}_{22}$ (bottom panel). We compare the unperturbed case with the perturbed one with $c=100M$ and $\epsilon=5\times10^{-4}$. While the Fourier transform of $\Gamma_{22}$ is stable against perturbations at all times, that of ${\cal R}_{22}$ acquires late-time corrections of ${\cal O}(\epsilon)$ that are reminiscent of echoes~\cite{Cardoso:2016oxy,Cardoso:2016rao,Cardoso:2017cqb}.}
	\label{fig:FTGammaR}
\end{figure}

This behavior is also reflected on the Fourier transforms of ${\cal R}_{lm}(\omega)$ and $\Gamma_{lm}(\omega)$, which is given in Fig.~\ref{fig:FTGammaR}, that compares the perturbed case with the unperturbed one. While the Fourier transform of $\Gamma_{lm}$ is insensitive to small perturbations, that of ${\cal R}_{lm}$ displays small variations at late-time which are reminiscent of echoes~\cite{Cardoso:2016oxy,Cardoso:2016rao,Cardoso:2017cqb}. While the latter remain of ${\cal O}(\epsilon)$, they slightly differ from the unperturbed case since the Fourier transform of ${\cal R}_{lm}$ vanishes at late times.

\begin{figure}[ht]
	\centering
	\includegraphics[width=0.48 \textwidth]{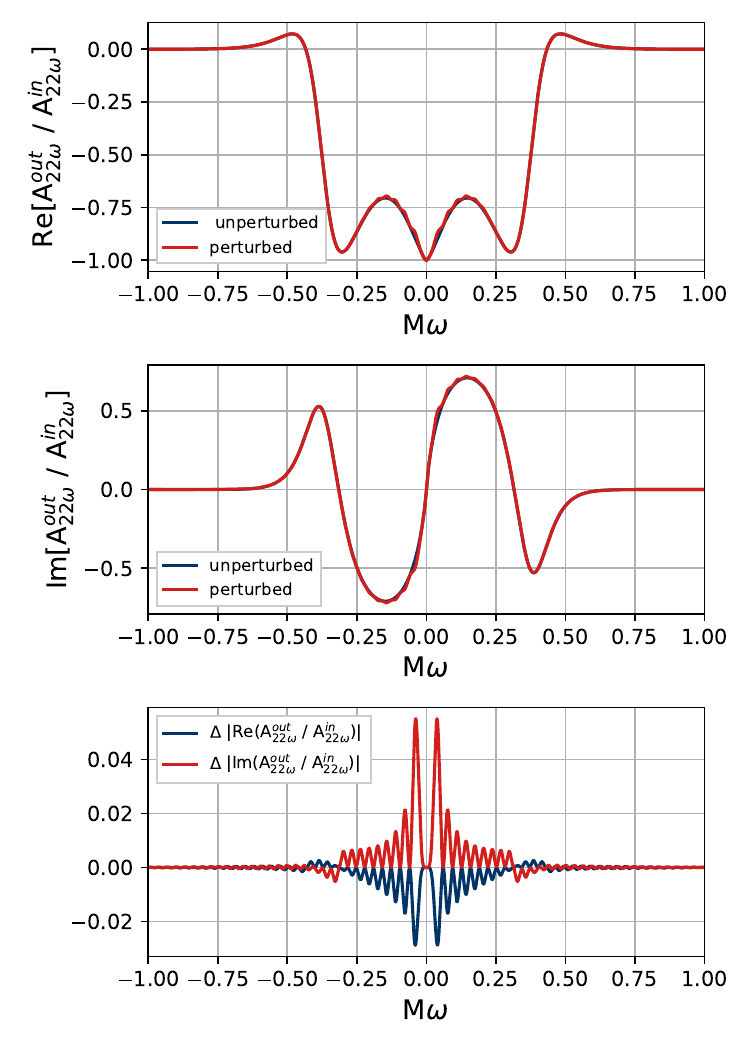}
	\caption{Real (top panel) and imaginary (middle panel) part of $A^{\rm out}_{22\omega}/A^{\rm in}_{22\omega}$ for an unperturbed BH and a perturbed one with $c=100M$ and $\epsilon=5 \times 10^{-4}$. The bottom panel shows the difference between the perturbed and unperturbed case for the absolute value of real and imaginary parts of the previous ratio. This highlights that these quantities are unstable at small frequencies, however their individual instabilities compensate each other such that the absolute value $|A^{\rm out}_{22\omega}/A^{\rm in}_{22\omega}|$ remains stable at low frequencies.}
	\label{fig:ReIm}
\end{figure}

The stability of ${\cal R}_{lm}$ and $\Gamma_{lm}$ is actually highly nontrivial. Indeed, considering the (complex) amplitudes of the incident and reflected wave, $A^{\rm in}_{lm\omega}$ and $A^{\rm out}_{lm\omega}$, we can separately study the stability of the real and imaginary part of the ratio $A^{\rm out}_{lm\omega}/A^{\rm in}_{lm\omega}$. While ${\cal R}_{lm}=|A^{\rm out}_{lm\omega}/A^{\rm in}_{lm\omega}|^2$ is stable at small frequencies, Fig.~\ref{fig:ReIm} shows that separately the real and imaginary parts are \emph{unstable} and acquire percent corrections at low frequencies even when $\epsilon=5\times10^{-4}$. These corrections compensate each other in the absolute value, yielding a stable reflectivity at low frequency. The same occurs for the real and imaginary parts of $1/A^{\rm in}_{lm\omega}$, which are separately unstable at small frequencies but their absolute value (yielding the GF) remains stable. 
The latter result is in agreement with the analysis of~\cite{Kyutoku:2022gbr}, which showed that the bump mostly destabilizes the phase shift of the reflected waves, rather than their amplitude.

\section{Ringdown reconstruction and QNM spectral instability}

\begin{figure}[ht]
	\centering
	\includegraphics[width=0.48 \textwidth]{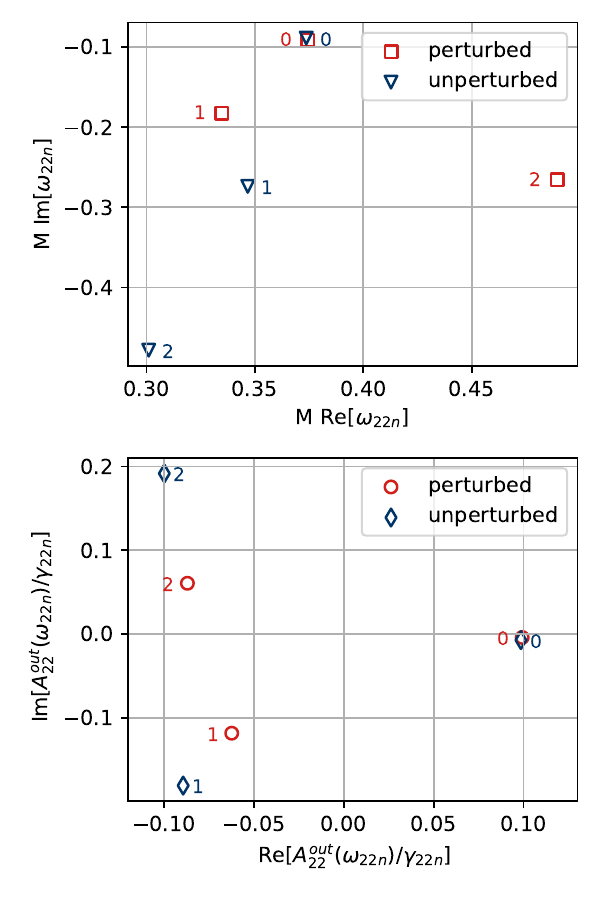}
	\caption{Comparison of perturbed and unperturbed Schwarzschild QNMs $\omega_{lmn}$ (top panel) and their corresponding excitation factors appearing in Eq. (6) in the main text (bottom panel) and used in the QNM decomposition of the time-domain signal shown in Fig. 4 in the main text. The perturbed case corresponds to the effective axial potential in Eq. (4) of the main text with $c=15M$ and $\epsilon=10^{-3}$.
	}
	\label{fig:perturbedunperturbed}
\end{figure} 

In the main text we showed how the time-domain signal at intermediate times (which is stable against small perturbations of the system) can be reconstructed by a superposition of either unperturbed or perturbed QNMs. This is highly nontrivial given that the QNMs are spectrally unstable, as shown in Fig.~\ref{fig:perturbedunperturbed} for the specific example of Fig. 4 in the main text, namely $c=15M$ and $\epsilon=10^{-3}$. As can be seen from the top panel of Fig.~\ref{fig:perturbedunperturbed}, the fundamental ($n=0$) QNM is approximately stable for this choice of $c$ and $\epsilon$, whereas the overtones ($n=1,2$) are unstable: when the effective potential is perturbed they acquire ${\cal O}(1)$ corrections. This is also reflected in the QNM excitation factors needed to reconstruct the time-domain signal. As shown in the bottom panel of Fig.~\ref{fig:perturbedunperturbed}, the excitation factors for $n=1,2$ in the perturbed case are significantly different than those in the unperturbed case, even for $\epsilon=10^{-3}$. It is therefore remarkable that, in both cases, the time-domain signal is recovered by the QNM decomposition (see Eq. (6) in the main text).

\begin{figure*}[ht]
	\centering
	\includegraphics[width=0.9\textwidth]{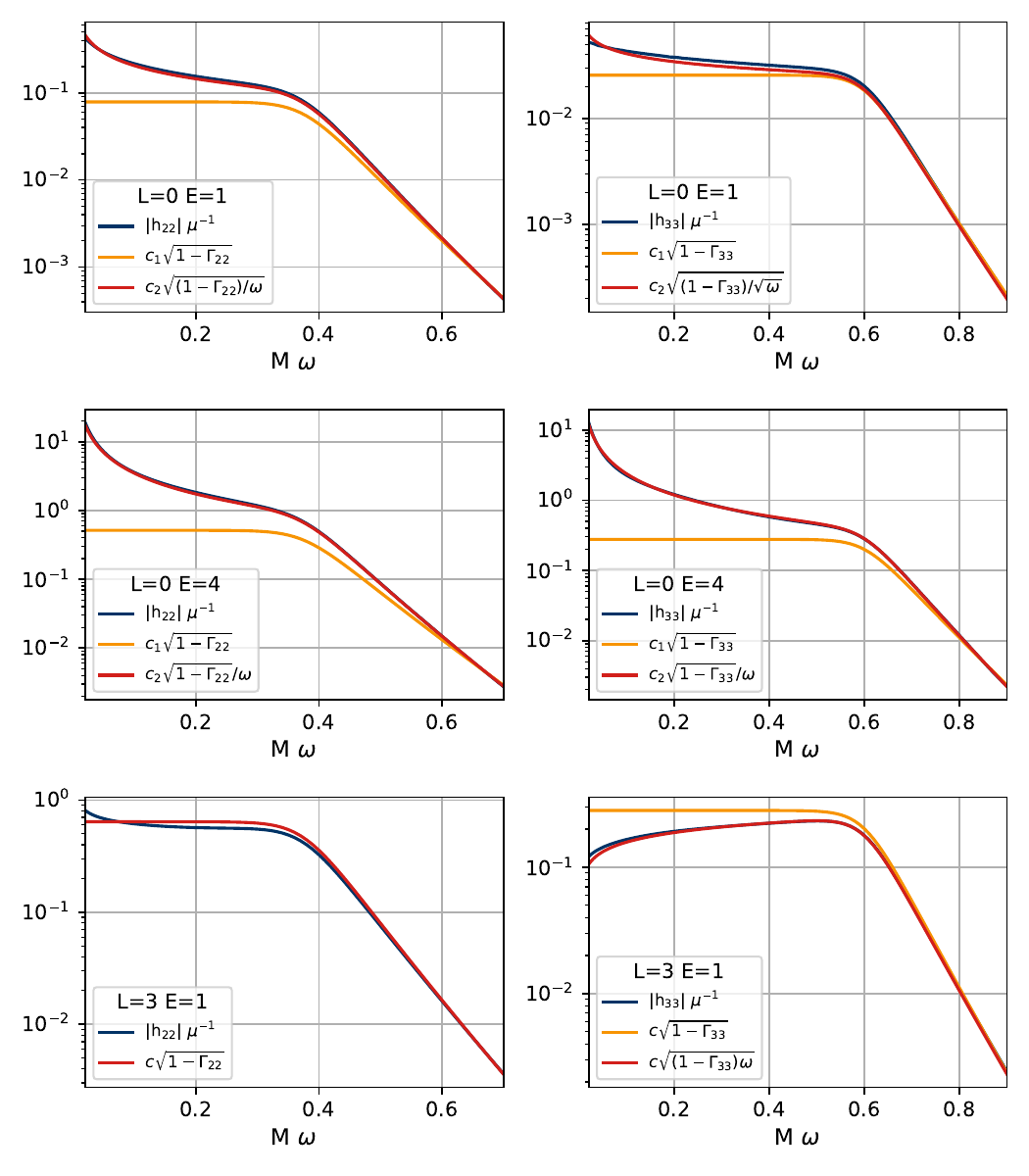}
	\caption{The spectral amplitudes for $l=2,3$ modes emitted by a point particle with mass $\mu$ and geodesic parameters $E$ and $L$ (see legends in each panels). Left panels refer to $|h_{22}|$ while right panels refer to $|h_{33}|$. For comparison, we also show a single-parameter model $\propto\sqrt{1-\Gamma_{lm}}/\omega^p$ for different values of $p$.}
	\label{fig:h22_differentEandL}
\end{figure*}

\section{GFs and ringdown spectral amplitudes}
In Fig.~\ref{fig:h22_differentEandL} we show a collection of results for the spectral amplitude $|h_{22}|$ and $|h_{33}|$ for a point-particle with different values  of $E$ and $L$, showing the universality of the behavior $|h_{lm}|\sim \sqrt{1-\Gamma_{lm}}$ at large frequencies. The spectral amplitude is computed as 
\begin{equation}
	h_{lm}(\omega)= H_{lm}(\omega) {}_{-2}{Y}_{lm}(\theta,\phi)\,,
\end{equation}
where ${}_{-2}{Y}_{lm}$ are the spin-weighted spherical harmonics (here and in the main text we assume $\theta=\pi/2$ and $m=0$ without loss of generality), and the amplitude is computed through the Green's function as
\begin{equation}\label{Xout}
	H_{lm}(\omega) = \frac{1}{16\imath\omega A_{lm\omega}^{\rm in}} \int_{-\infty}^{+\infty}dr_*  X_{lm\omega}^{\rm hom} S_{lm\omega}\,,
\end{equation}
where $X_{lm\omega}^{\rm hom}$ is the solution of the homogeneous equation associated with Eq. (1) with ingoing boundary conditions at the horizon. The source term can be found in \cite{Oohara:1984ck} and, for $l=m$ modes, it excites the polar sector.

The two polarizations of the time-domain signal read~\cite{Oohara:1984ck}
\begin{equation}
	h_{+}+\imath h_{\times}= \frac{1}{r} \sum_{lm} \int_{-\infty}^\infty d\omega e^{\imath \omega (r_* -t)}H_{lm}(\omega){}_{-2}{Y}_{lm}(\theta,\phi)\,.
\end{equation}
Numerical computations have been performed using the methods outlined in \cite{Silva:2023cer} in order to improve the convergence of the integral in Eq.~\eqref{Xout}.

In Fig.~\ref{fig:h22_differentEandL}, we also compare the spectral amplitude with a model $\propto\sqrt{1-\Gamma_{lm}}/\omega^p$ for different values of $p$. It is remarkable that the spectral amplitude is very well modelled for \emph{any frequency by a single parameter} function where the value of $p$ depends on $E$, $L$, and $l$. It would be intriguing if a similar universality occurs also beyond the point-particle regime~\cite{Okabayashi:2024qbz}. We expect that, in general, $|h_{lm}(\omega)|\propto \sqrt{1-\Gamma_{lm}}f(\omega)$, with $f(\omega)$ interpolating between unity at large frequency and a source-dependent (but possibly easy to model) behavior at low frequency. We postpone a more detailed analysis on this aspect to future work.

\begin{figure}
	\centering
	\includegraphics[width=0.48 \textwidth]{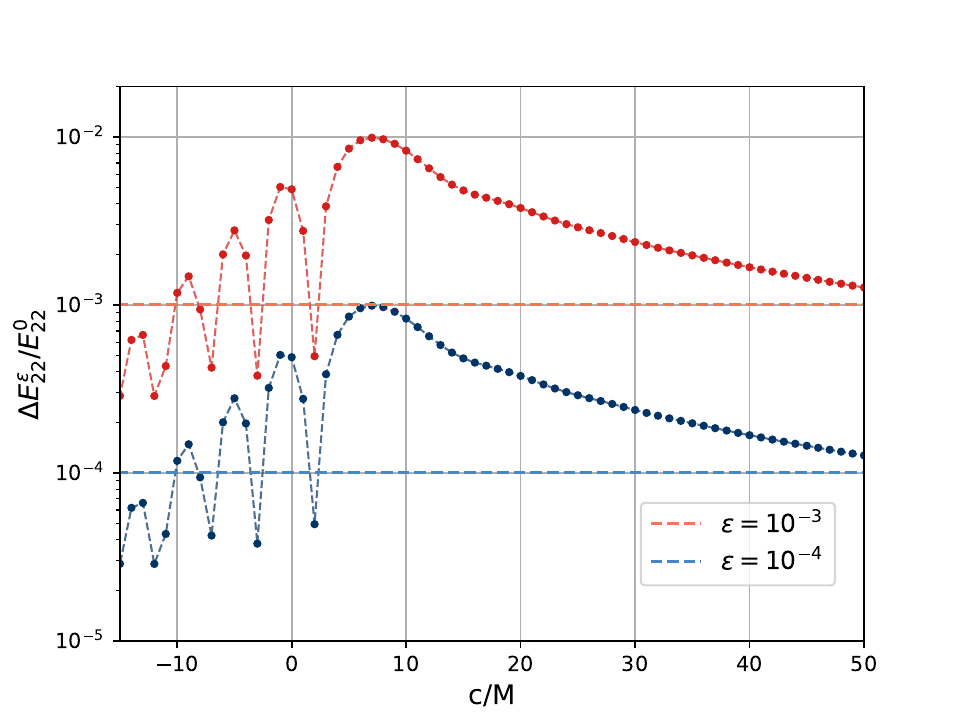}
	\caption{Relative difference of the total emitted energy in $l=2$ modes by a point particle plunging into a Schwarzschild BH with $E=1$ and $L=0$ as a function of the bump location $c$ for $\epsilon= 10^{-3}$ and $\epsilon= 10^{-4}$. Different values of $E$ and $L$ show the same stability behavior.}
	\label{fig:energyperturbation}
\end{figure}

Finally, in Fig.~\ref{fig:energyperturbation} we show the stability of the total energy for a given $(l,m)$ mode,
\begin{equation}
	E_{lm}=\int_0^{+\infty} d\omega \,\frac{dE_{lm}}{d\omega} = \frac{1}{4}\int_0^{+\infty} d\omega \,\omega^2 \Big|H_{lm\omega}\Big|^2\,.
\end{equation}
Note that, at least at large frequency, also the energy flux displays a universal behavior depending only on the GF, $dE_{lm}/d\omega\sim 1-\Gamma_{lm}$, since this quantity is proportional to $|h_{lm}|^2$.

\bibliography{biblio}

\end{document}